\documentclass[%
 reprint,
 amsmath,amssymb,
 aps,nofootinbib,   
]{revtex4-1}
\usepackage{bm}% bold math
\PassOptionsToPackage{linktocpage}{hyperref}
\usepackage[hyperindex,breaklinks]{hyperref}
\usepackage{enumitem}
\usepackage{slashed}

\renewcommand{\theta}{\vartheta}

\newcommand{\bra}[1]{\ensuremath{\left< #1\,\right|}}
\newcommand{\ket}[1]{\ensuremath{\left|\, #1\right>}}

\usepackage{array}
\usepackage{mathtools}

\usepackage{etoolbox}

\begin{document}

\title{ Area Law Saturation of Entropy Bound from Perturbative Unitarity in Renormalizable Theories}

\author{Gia Dvali} 
%$^{1,2,3}$}
%\email{Georgi.Dvali@physik.uni-muenchen.de}
\affiliation{%
%$^1$ 
Arnold Sommerfeld Center, Ludwig-Maximilians-Universit\"at, Theresienstra{\ss}e 37, 80333 M\"unchen, Germany, 
}%
 \affiliation{%
%$^2$ 
Max-Planck-Institut f\"ur Physik, F\"ohringer Ring 6, 80805 M\"unchen, Germany
}%
 \affiliation{%
%$^3$ 
Center for Cosmology and Particle Physics, Department of Physics, New York University, 726 Broadway, New York, NY 10003, USA
}%

\date{\today}

\begin{abstract}  
We study the quantum information storage capacity of 
solitons and baryons in renormalizable quantum field theories 
that do not include gravity.   We observe that 
a 't Hooft-Polyakov magnetic monopole saturates 
the Bekenstein bound on information  when the theory saturates the bound on perturbative unitarity.  In this very limit the 
monopole entropy assumes the form of an area-law, strikingly similar to a black hole entropy in gravity.  The phenomenon 
appears universal and takes place for other solitons and non-perturbative objects. We observe the same behaviour of entropy of a baryon in QCD with large number of colors.  
  These observations indicate that the area-law form of the entropy bound  extends  beyond gravity and is deeply rooted in concepts of  
weak coupling  and perturbative unitarity.   
 One provoked idea is that confinement 
  in QCD may be understood as a prevention mechanism against violation of Bekenstein entropy bound by colored states.   
\end{abstract}

\maketitle
\section{Introduction}

The celebrated Bekenstein entropy bound 
\cite{BekBound} 
 tells us that the maximal amount of information stored in a system of energy $M$ and a size $R$ is given by (we drop the order-one numerical factors for simplicity)  
\begin{equation} \label{Bek1}
    S_{\max} =  MR \, . 
\end{equation}
This bound is {\it universal}  and is independent of the existence of gravity
(see also, \cite{BrBound}). 
However, in gravity, where it is saturated by a black hole, the bound
acquires a very interesting form.  
Mysteriously, this saturation comes in form of an area-low
\cite{BekE}. That is,  
the Bekenstein-Hawking entropy of a  black hole is given by its area
in units of the fundamental scale, the Planck mass $M_P$: 
\begin{equation} \label{Bek2}
      S_{\max} = MR = (RM_P)^2 \, . 
\end{equation}
 Understanding the origin of this relation in the context of 
 quantum gravity has been a subject of substantial effort. In particular, it has been derived in a string-theoretic construction of Strominger and Vafa \cite{SV}. Another framework for addressing this question in quantum gravity is provided by  AdS/CFT correspondence \cite{ADS1}. \\

 In the present paper we would like to ask a very different 
question.  Does the area-law form of the entropy bound extend beyond 
gravity and if yes what is its underlying meaning? \\
 
 We must comment that the above question was already asked in \cite{giaArea} where an explicit toy model with area-law entropy was 
 constructed in form of a  non-relativistic
 quantum-critical system.  
  However, the questions of saturation of the entropy bound and of relativistic completion of the model were not discussed. \\

In the present paper we would like to take a different approach and focus exclusively on  
renormalizable quantum field theories within the regime of 
weak coupling and perturbative unitarity. 
Such are gauge theories in Higgs and confining phases respectively.  
   These theories obviously do not contain black holes in their spectra. 
   However, they contain solitons and baryons.  By studying the quantum information storage capacity 
of solitons  and baryons we observe the following: 

\begin{itemize}
  \item The soliton (baryon) entropy saturates the Bekenstein bound on information storage when the theory saturates the bound on perturbative unitarity. 
  \item  Simultaneously, the soliton (baryon) entropy assumes the form of an area-law, 
  strikingly similar to the black hole relation (\ref{Bek2}). 
  
  \item In this limit the entropy also becomes equal to the inverse of the perturbative quantum coupling, 
  \begin{equation}
       S_{max} \,  = \, {1 \over g^2} \, = \, {\rm Area}. 
       \end{equation}   
\end{itemize} 
 We shall demonstrate the above behaviour on two examples. One is a  't Hooft-Polyakov magnetic monopole  and another is a baryon  
in QCD with large number of colors. 
 However, the phenomenon appears universal and holds for other theories with non-perturbative objects. 
  This observation indicates that the saturation of the information storage capacity and its relation to the area law goes well beyond gravity and 
  is deeply rooted in notions such as weak coupling and perturbative unitarity.  \\
  
  Viewing the states of quantum field theory from the point of view of 
  their information storage capacity can open up some new perspectives.  
 For example, since the Bekenstein entropy bound is fully non-perturbative and independent of the particular choice of the degrees of freedom, it may provide a powerful tool for probing non-perturbative regimes in quantum field theory.  \\

  One of the ideas provoked by the above observations is 
 a possible connection between the Bekenstein bound and the phenomenon  of confinement  in QCD.  We shall hypothesize and ask whether the confinement can be viewed 
 as a preventive mechanism by  which the theory avoids a potential violation of the Bekenstein bound.  Such violations could have been triggered by highly degenerate colored states that would be permitted to exist in an unconfined theory.

 \section{ Monopole} 
 We shall start with a simplest model that contains a
 't Hooft-Polyakov monopole \cite{Monopole} in its spectrum. This is a theory 
 with a gauged $SO(3)$ symmetry in the Higgs phase. This phase is  achieved due to a non-zero vacuum expectation  value (VEV) of a scalar field  $\Phi^a$ 
   in the triplet representation of the group, where 
   $a = 1,2,3$ is  an $S0(3)$-index.  The Lagrangian
   has the following form:  
     \begin{eqnarray}   \label{Lag1} 
  &&  L =  {1 \over 2} D_{\mu}\Phi^a D^{\mu}\Phi^a
  - {1 \over 4} F_{\mu\nu}^aF^{\mu\nu a}
     -  \\ \nonumber
   && - {h^2 \over 4} (\Phi^a\Phi^a - v^2)^2 \,,     
 \end{eqnarray}
  where as usual $D_{\mu}\Phi^a \equiv \partial_{\mu}\Phi^a
  + e \epsilon^{abc} A_{\mu}^b\Phi^c, ~~
 F_{\mu\nu}^a \equiv \partial_{\mu}A_{\nu} ^a -\partial_{\nu}A_{\mu} ^a
  + e \epsilon^{abc} A_{\mu}^bA_{\nu}^c$. The parameters 
  $e$ and $h$ are dimensionless coupling constants, whereas 
  $v$ represents a fundamental energy scale of the theory.  \\
   
 As it is well known, this theory splits into several super-selection sectors 
 determined by the value of the magnetic charge.   
 In the topologically trivial sector the VEV 
 of the Higgs field can be chosen as
 \begin{equation} \label{Higgsvacuum}
   \Phi^a  = \delta^{a3} v  \, .
  \end{equation} 
  The $SO(3)$ gauge group is Higgsed down to its Abelian 
  $U(1)$-subgroup.
  The corresponding gauge boson (``photon") $A_{\mu}^3$ remains massless whereas the two other gauge bosons $A_{\mu}^{1,2}$ gain masses equal to $m_v = ev$. The longitudinal components 
of these massive gauge fields come from two Goldstone 
degrees of freedom that parameterize all possible rotations
of the VEV (\ref{Higgsvacuum}) in the internal space without changing 
its length.  The third scalar (the Higgs boson),  that described the fluctuations of the absolute value has a mass $m_h = hv$. \\

 We must note that in the absence of the Higgs boson,  the longitudinal polarizations (equivalently, the Goldstone bosons) would violate perturbative unitarity above the scale $v$. In this sense the parameter $v$ sets an upper 
 bound on the scale of UV-completion of the low energy theory of massive 
 gauge bosons (or Goldstone bosons). \\

  We focus our attention on topologically non-trivial vacua 
 the simplest of which carries an unit  magnetic charge.  
   This vacuum is described by a topologically-stable 
 and  spherically-symmetric solution of the following form
 \cite{Monopole}
\begin{equation}\label{Mon}  
 \Phi^a = {x^a \over r} vH(r), ~~ A_{\mu}^a = {1 \over er} 
 \epsilon^{0a\mu\nu} {x_{\nu} \over r} F(r)\,,
 \end{equation} 
 where $r$ is the radial coordinate.  
The asymptotic values of the two functions are 
$H(0)=F(0)= 0,\, ~H(\infty) = F(\infty) = 1$. 
This solution describes a 't Hooft-Polyakov monopole \cite{Monopole}. 
Notice, the monopole configuration is spherically-symmetric 
in generalized sense since it is invariant under the combined 
rotations in coordinate and internal spaces. \\

 The size of the monopole  core $R_{mon}$ is determined by the size of the region  where $H(r)$ and  $F(r)$ deviate from one significantly and is given by 
 the Compton wavelength of the gauge boson 
 $R_{mon} = m_v^{-1} = (ev)^{-1}$. 
 The mass of the monopole is 
 \begin{equation} \label{mass}
 M_{mon} \sim {m_v \over e^2} \,
 \end{equation} 
 up to an order-one coefficient that becomes exactly $4\pi$ in the BPS limit 
 of $h=0$. 
  The magnetic charge of the monopole 
 is $q_{m} = {1 \over e}$ in agreement with Dirac's charge quantization 
 rule. \\
 
We are interested in the spectrum of quantum excitations
in the monopole background.  
   In the above simplest model the  monopole carries only few gapless excitations in its spectrum.   Such are the translation moduli that represent Goldstone bosons of broken translational invariance.
 Correspondingly,  at least within semi-classical description, there exists  no richness of gapless modes
for the efficient storage of quantum information.  This can be quantified by saying that in the above minimal model the monopole carries
a very little micro-state entropy. \\   

  We shall now propose the two distinct mechanisms that shall  allow us to    increase the micro-state entropy of the monopole in a controllable way.    
 We shall then study what happens when the monopole saturates the
 Bekenstein bound  (\ref{Bek1}) on the information storage capacity. 
 
  \section{Goldstone zero modes} 
  
 In the first scenario the monopole entropy is increased due 
 to appearance of  localized bosonic gapless modes. These modes 
 originate from the Goldstone bosons of a global 
 $SO(N)$ symmetry that is spontaneously broken within the monopole core 
 but is restored outside. Consequently, the  Goldstone modes are localized in the monopole world-volume.  In order to achieve this picture, we couple 
 the monopole Higgs field $\Phi^a$ to a new scalar field 
 $\sigma_{\alpha}$ which transforms under some large global 
 ``flavor"
 symmetry group.  For definiteness, we choose the global flavor symmetry to be $SO(N)$ with 
 $N$ arbitrarily large. We choose  $\sigma_{\alpha}, \alpha = 1,2,...N$ to be a real scalar field transforming as $N$-dimensional vector representation
 of $SO(N)$. The generalizations of our idea to other groups and representations is trivial.     
 Next, we arrange the parameters in such a way that  
$\sigma_{\alpha}$ acquires a vacuum expectation value 
within the monopole core.   \\

 The condensation of 
 $\sigma_{\alpha}$ inside the monopole is inspired by condensation of a charged field on a cosmic string in 
the model of a superconducting cosmic string proposed by Witten
\cite{Witten}. However, it is crucial for us that symmetry is global and $N$ is large. 
  In this sense our construction is spiritually linked to the one of
  \cite{texture}. There, as a result of analogous condensation, a set of  non-Abelian gapless Goldstone bosons  becomes localized inside a topological defect.  \\
 
 In order to achieve our goal we add the following terms to the 
 Lagrangian 
   \begin{eqnarray}   \label{sigma} 
    &&  L_{\sigma} = {1 \over 2} \partial_{\mu}\sigma_{\alpha} 
  \partial^{\mu}\sigma_{\alpha} \, - \, 
  {1\over 4} g_{\sigma}^2 ( \sigma_{\alpha}\sigma_{\alpha})^2 
     -  \\ \nonumber
   && - {1 \over 2}(
   g^2 \Phi^a\Phi^a  - m^2)  
   ( \sigma_{\alpha}\sigma_{\alpha}) \,,     
 \end{eqnarray}  
  The last term  consists of the ``bare" mass 
  $-m^2 < 0$ and its interaction with the monopole field 
 controlled by the coupling $g^2 > 0$.  The combination of the two contributions evaluated on the monopole 
 background gives an effective $r$-dependent 
mass term for the $\sigma$ field which is given by 
 $m^2(r) =  g^2 v^2 H(r) - m^2$. 
 If $m^2(\infty) =  g^2 v^2 - m^2 > 0$, the VEV of the $\sigma$-field vanishes
 in the  asymptotic vacuum and there exist no gapless excitations at infinity.  
 However, the effective mass term is negative in the monopole core. 
 We wish to chose parameters in such  a way that
 it is energetically favorable for $\sigma$ to condense inside the monopole. 
 The fact that such a choice is possible follows from the following 
 argument.  Let us first set  $g_{\sigma} =0$ and $g^2 v^2 - m^2 = 0$.
 Then the linear perturbations of the $\sigma$-field see the monopole as 
 a semi-negative definite potential well of the depth
 $ - (gv)^2$ and the radius $m_v^{-1}$. Inside the well the 
 perturbations around $\sigma =0$ are tachyonic with a negative 
 mass-square given by  $ - (gv)^2$. However, in order for the tachyons to condense, the well must be sufficiently wide. Namely, the radius must 
 be larger than the Compton wavelength of the tachyon. This gives a condition
 \begin{equation}\label{eg1}
 g/e \gtrsim 1\,. 
 \end{equation}
   In order to make this more obvious, consider a  spherically symmetric 
 configuration
 $\sigma_{\alpha} (r) = \sigma(r)\delta_{\alpha1} $
  in the monoploe background. 
 The energy of it is given by 
 the following functional
 \begin{equation} \label{energy}  
 E = 2\pi \int_0^{\infty} r^2dr \left ((d_r\sigma(r))^2 -  m^2(r)\sigma(r)^2
 \right ) \, .
 \end{equation} 
 Assume that $\sigma(r)$ is some smooth probe function 
 with a localization radius $R$ and an amplitude $A$ (for example, $\sigma(r)=A{\rm e}^{-r^2/R^2}$). 
  For $R \gtrsim  m_v^{-1}$, the energy of such a configuration can be estimated as 
 \begin{equation} \label{energy1}  
  E \sim A^2 \left (R -  {g^2 \over e^2} m_v^{-1} \right )\, .
 \end{equation}      
  Thus, having $A\neq 0$ is energetically favourable as long as,  
 \begin{equation} \label{local}
    R \lesssim  {g^2 \over e^2} m_v^{-1} \, , 
  \end{equation} 
 which also sets the bound on the localization width of the $\sigma$-condensate.    
Obviously, the tendency that $\sigma$ condenses 
   inside the monopole is maintained also for 
 $g_{\sigma}^2 > 0$ and $g^2 v^2 - m^2 > 0$ as long as 
 these parameters are not too large.  
 \\
 
 A straightforward detailed estimate of the energy balance shows that the 
 energy is minimized by the following values of the localization 
 width of the $\sigma$-condensate and its value in the center of the monopole 
    \begin{equation} \label{scales1}
    R \sim (ev)^{-1},\, ~  \, \sigma(0) \sim {g \over g_{\sigma}} v\,.
  \end{equation} 
   The requirement that the back-reaction from the condensate 
  $\sigma(r)$ to the monopole solution (\ref{Mon})  is weak, implies 
     \begin{equation} \label{scales2}
    g^2  \lesssim g_{\sigma} e\,.
  \end{equation} 
In this case the correction to the monopole mass due to the 
$\sigma$-condensate is small and the total energy of the configuration is well described  by (\ref{mass}).  \\
 
  We thus see that there exists a choice of parameters for which 
  the ground-state of the system is described by a function 
  $\sigma_{\alpha} (r)$ that has a non-zero expectation value in the monopole core. Thus, $SO(N)$-symmetry  is broken spontaneously
  down to $SO(N-1)$ within the monopole.    
  We are interested in $N-1$ Goldstone bosons that originate from this breaking. Namely, we are interested in the zero angular-momentum excitations of the Goldstone fields. They represent the gapless modes
 that account for the micro-state degeneracy of the monopole vacuum. \\
  
  In the classical limit, these modes  parameterize all possible ground-state 
 configurations of the form,  
  \begin{equation}\label{gold}  
  \sigma_{\alpha} = {\sigma(r) \over \sqrt{N_{\sigma}}}  a_{\alpha}\,,
  \end{equation} 
  subject to the constraint 
 \begin{equation}\label{conN}  
 \sum_{\alpha=1}^N a_{\alpha}^2 = N_{\sigma} \,, 
  \end{equation} 
where, 
\begin{equation}\label{Nsigma}  
  N_{\sigma} \sim \int \sigma(r)^2m(r) \sim {g^3 \over g_{\sigma}^2e^3}  \, . \end{equation} 
  We can think of $N_{\sigma}$ as the vacuum average occupation number of quanta in the $\sigma$-condensate (or a coherent state)  formed due to an attractive potential energy of the monopole. \\
 
In quantum theory $a_{\alpha}$ must be interpreted as the 
expectation values of the corresponding
creation/annihilation 
 operators that satisfy the usual algebra, 
 $[\hat{a}_{\alpha}, \hat{a}_{\beta}^{\dagger}] = \delta_{\alpha\beta}$. 
 Their effective Hamiltonian describing the structure of the vacuum 
 has a very simple form, 
 \begin{equation}\label{gold}  
  \hat{H}  = X \left(\sum_{\alpha = 1}^N \hat{a}_{\alpha}^{\dagger} \hat{a}_{\alpha} - N_{\sigma}\right ) \, ,   
  \end{equation} 
  where $X$ is a Lagrange multiplier that imposes the constraint 
  (\ref{conN}). The degenerate microstates are 
 thus given by all possible number eigenstates of the form 
 \begin{equation}\label{states} 
 \ket{n_1,...,n_{N}} \, ,
 \end{equation}
 with the constraint $\sum  n_{\alpha} = N_{\sigma}$,  where  
 $n_{\alpha}$ are the eigenvalues of corresponding number operators
$\hat{n}_{\alpha} \equiv \hat{a}_{\alpha}^{\dagger}\hat{a}_{\alpha}$. \\
 
 The total number of such states is given by the binomial coefficient 
  \begin{equation}\label{Nstates} 
 n_{st} =  \begin{pmatrix}
    N_{\sigma} + N -1   \\
     N_{\sigma}   
\end{pmatrix} \,,
  \end{equation}
  which  for  $N \sim N_{\sigma}$ scales exponentially with $N$. 
  For example, for $N = N_{\sigma}$ using Stirling's approximation we have  $n_{st} \sim {2^{2N} \over \sqrt{\pi N}}$.
 The corresponding micro-state entropy (up to an order-one logarithmic factor) scales as, 
 \begin{equation}\label{Sstate} 
 S_{mon} = \ln (n_{st}) \sim N \, .
  \end{equation}
  
An alternative way for  obtaining (\ref{Nstates}) is to notice that
the localized $\sigma$-condensate is a symmetric $N_{\sigma}$-particle state. Each constituent of the condensate can assume an arbitrary value of the $SO(N)$-index $\alpha = 1,...,N$.  
 Thus, the ground-state wave-function represents a 
$N_{\sigma}$-particle wave-function which transforms as 
a symmetric tensor of rank-$N_{\sigma}$ under the flavor 
group $SO(N)$. Its components (up to obvious normalization factors) can be written  as, 
   \begin{equation}\label{Nstates1} 
 \Psi_{\alpha_1,...,\alpha_{N_{\sigma}}} = 
 \sum_n\bra{n_1,...,n_N} \prod_{j=1}^{N_{\sigma}} \hat{a}_{\alpha_j}^{\dagger}
 \ket{0}_{\sigma} \, ,  
  \end{equation}
where indexes $j$ and $\alpha_j$  take values $j=1,...,N_{\sigma}$ and 
$\alpha_j = 1,...,N$ respectively. 
In this expression $\ket{0}_{\sigma}$ is the vacuum on which the condensate 
is built. The equation (\ref{Nstates}) is simply a dimensionality 
of such a tensor.\\

 Here emerges a curious analogy which shall become more transparent later. 
 Notice, if  we think of $N_{\sigma}$ as the number of ``colors", 
 then the monopole with $\sigma$-condensate can be interpreted  
 as a ``baryon"-like  state with $SO(N)$ playing the role of the 
 quark flavor group of QCD.  
 As we shall see, this similarity is not an accident. \\

 Notice, if the spontaneous breaking of $SO(N)$ symmetry would  take place in 
 the entire space, the above gapless modes would correspond 
 to infinite-wavelength excitations of the Goldstone modes and would account for the vacuum degeneracy.  However, in the present case 
 the condensate is confined within the monopole and so are  
the gapless modes. They therefore belong to the world-volume theory 
of the monopole. Consequently, the degenerate states created by excitations of these gapless modes 
contribute into the micro-state count of the monopole. In other words, in quantum theory the lowest energy excitations of  localized Goldstone bosons 
can be viewed as gapless qudits that encode 
quantum information.  The measure of this information is given 
by the micro-state entropy (\ref{Sstate}).\\

 Any potential doubt whether (\ref{states}) must be counted as distinct micro-states can be eliminated  by means of a {\it soft}  explicit breaking of 
   the global flavor $SO(N)$-symmetry and later taking the exact symmetry limit. 
 The sole effect of such soft breaking is a generation of tiny mass gaps for the Goldstone modes. This results into a level splitting and a slight lift of the  degeneracy of the micro-states.  The limit of exact symmetry is smooth and 
confirms the necessity of inclusion of all the states (\ref{states})   
into the entropy count. \\

  We now wish to  evaluate a maximal entropy that can be carried by the monopole.   For this, we must take into account the various restrictions on $N$ and 
  $N_{\sigma}$.  First, the perturbative unitarity puts the following bound, 
  \begin{equation} \label{Nbound}
    g^2N \lesssim 1,\, ~  \, g_{\sigma}^2N \lesssim 1\, .
  \end{equation} 
  Taking the saturation point of this bound and using the relations
   (\ref{eg1}), (\ref{scales2}) and (\ref{Nsigma}), we arrive to the following 
  limiting expressions, 
   \begin{equation} \label{Nsaturation}
    {1\over g^2} \sim {1 \over e^2} \sim {1 \over g_{\sigma}^2} \sim 
    N\, \sim N_{\sigma} .
  \end{equation} 
  Now, from the first relation in (\ref{scales1}) and the 
  expression for the monopole mass (\ref{mass})
  it is clear that the maximal entropy of the monopole can be written as 
  \begin{equation} \label{Nsaturation}
    S_{mon}  = N = (R_{mon}v)^2\ =  M_{mon}R_{mon}     
  \end{equation} 
The first part of this equality tells us that at the point 
of saturation of the unitarity bound the monopole entropy $N$
becomes equal to its area in units of the fundamental scale $v$. 
 At the same time, the last equality in (\ref{Nsaturation})
tells us that this area-low form of monopole entropy simultaneously saturates the Bekenstein bound (\ref{Bek1}) on information storage!

   \section{Fermion Zero Modes} 
   
    An alternative mechanism that endows a monopole with  
a controllable  micro-state entropy works through the localization of fermionic zero modes. 
In order to accomplish this, we need to
ensure that the Higgs field $\Phi^a$ gives masses to 
some fermion species though the Yukawa couplings.  It is well known that in such a case the fermions deposit some zero modes in the monopole 
core \cite{JR}. This phenomenon is a consequence of a general index theorem \cite{index}. \\
   
In order to achieve this, we choose the fermion content in form of 
 two real Majorana fermions $\psi_{\alpha}^a, 
 \lambda_{\alpha}^a$
  transforming as triplets under the gauge $SO(3)$ group and in the same time forming $N$-dimensional vector representations of a global 
  $SO(N)$-flavor symmetry group.  The labels $a=1,2,3$ and $\alpha = 1,2,.....N$ are 
  $SO(3)$ and $SO(N)$ indexes respectively.  The fermionic part of the Lagrangian has the following form: 
 \begin{eqnarray}   \label{sigma} 
    &&  L_{\sigma} = {1 \over 2} \bar{\psi}_{\alpha}^a
    \gamma^{\mu}D_{\mu}
   \psi_{\alpha}^a \, + \,  
   {1 \over 2} \bar{\lambda}_{\alpha}^a
    \gamma^{\mu}D_{\mu}
   \lambda_{\alpha}^a 
       -  \\ \nonumber
   && - 
   g \epsilon^{abc}\Phi^a  \bar{\psi}_{\alpha}^b \lambda_{\alpha}^c \,,     
 \end{eqnarray} 
 where $g$ is  a dimensionless coupling constant. 
 Notice, we use real  $\gamma^{\mu}$-matrixes. \\  
  
 In the topologically-trivial vacuum (\ref{Higgsvacuum}) the 
 fermions $\psi_{\alpha}^1, \lambda_{\alpha}^2$ and 
  $\psi_{\alpha}^2, \lambda_{\alpha}^1$ pair up and form
  the two Dirac fermions with the masses $m_f = gv$ per each flavor $\alpha$. 
  The third pair  $\psi_{\alpha}^3, \lambda_{\alpha}^3$ remains massless
  for all values of $\alpha$.  \\  
    
   In the monopole background a dramatic novelty takes place in form 
   of the appearance of fermionic zero modes localized 
   within the monopole core.  This phenomenon takes place for arbitrary 
   values of the couplings as long as $g \neq 0$. 
   For example, for $h=0$ and $e=g$ the zero-frequency solution 
  (up to an over-all finite normalization constant) takes the form 
  \begin{eqnarray}   \label{sigma} 
    &&  \lambda_{\alpha}^a = {1\over 2} F_{\mu\nu}^a\sigma^{\mu\nu} \epsilon_{\alpha}  
         \\ \nonumber
   &&
   \psi_{\alpha}^a = \gamma^{\mu} D_{\mu} \Phi^a \epsilon_{\alpha} \,,     
 \end{eqnarray}  
   where $\epsilon_{\alpha}$ ($\alpha = 1,2,...N$) are the constant spinors
   and all the bosonic fields take the form  (\ref{Mon}). \\
   
  The  appearance of $\sim N$ fermionic zero modes generates 
 $n_{st} \sim 2^{N}$ degenerate micro-states in the monopole spectrum. Correspondingly the monopole acquires a  micro-state entropy 
that similarly to (\ref{Sstate}) scales as $S_{mon} \sim N$. \\

Let us establish the relation between the monopole entropy
 and the size and the mass of the system in the limit when the theory saturates the bond on perturbative unitarity. 
 The later bound takes the form similar to (\ref{Nbound}): 
  \begin{equation} \label{NboundF}
    g^2N \lesssim 1,\, ~  \, e^2N \lesssim 1\, .
  \end{equation} 
 The inessential difference from the Goldstone case is that the gauge coupling $e$ is directly bounded because the fermions transform  
  under the $SO(3)$ gauge group.   Note,  the quantity $e^2N$ plays the role analogous 
  to the 't Hooft coupling  in planar QCD \cite{planar} with the difference that 
  in the present case $N$ is counting the number of flavors rather than colors.  \\
   
 Now,  taking into account that the size of the system (i.e., the localization radius of the fermionic zero modes) is given by the size of the monopole 
  $R_{mon}= (ev)^{-1}$ and that the energy is given by its mass (\ref{mass}), it is clear that the maximal entropy of the monopole  
 satisfies the relation (\ref{Nsaturation}).  Thus, 
we observe the same behaviour of the monopole 
entropy as in the previous example of localized Goldstone modes. 
At the limit of perturbative unitarity (\ref{NboundF}) 
the entropy of the magnetic monopole saturates the Bekenstein bound and simultaneously assumes the area 
 law. 
 
 \section{Baryons} 
 
  The  phenomenon that we have observed for a magnetic monopole 
  appears to be universal and takes place also in other systems.   Another important example 
  is given by a baryon in QCD. We consider QCD with 
  $SU(N_c)$ color group, a gauge coupling $g$ and $N$ flavors of quarks. Each flavor consists of a pair of left-handed and right-handed Weyl 
 fermions in the fundamental representations of $SU(N_c)$.    
 Therefore, if quarks are massless, the Lagrangian exhibits a global flavor  symmetry 
$SU(N)_L \otimes SU(N)_R \times U(1)_A\times U(1)_B$, 
where $U(1)_B$ is the baryon number and $U(1)_A$ is the anomalous axial 
symmetry.
  
  We shall work in 't Hooft's  planar picture \cite{planar} in which $N_c$ is arbitrarily large while $g^2N_c$ is finite.
 The QCD scale $\Lambda$ is also finite. 
 The non-Abelian global flavor  symmetry of quarks is spontaneously broken down to a diagonal 
 subgroup $SU(N)_F$ by the quark condensate. This breaking results 
 into massless Goldstone bosons, the pions.  The additional 
 would-be Goldstone boson of broken anomalous $U(1)_A$-symmetry, the analog of  $\eta'$-meson,  is getting mass from instantons \cite{tHooftA,tHooftDet} and $1/N_c$ effects \cite{WV}. \\ 
 
  The low energy effective 
 theory for pions is a chiral Lagrangian. 
 An important  scale in the problem is the pion decay constant $f_{\pi} \sim \sqrt{N} \Lambda$.  
 This is the scale above which, without a proper UV-completion, 
 the pion-pion interaction would become strong and would violate unitarity. 
 In this respect $f_{\pi}$  plays the role 
 analogous to the Higgs VEV  $v$ in the above monopole example. It also plays the role analogous to $M_P$ in gravity. \\
 
 The mass and the size of a baryon in this limit are given by \cite{WittenN}, 
 \begin{equation} \label{Bmass} 
 M_B = N_c\Lambda \,,~ \,R_B = \Lambda^{-1} \, .
 \end{equation} 
  There exist two alternative descriptions of baryons. One 
  is a bound-state of $N_c$ quarks.  Another is a soliton of pions, a so-called skyrmion \cite{skyrme}. Witten has shown  \cite{WittenS} that 
  skyrmion/baryon correspondence is exact in 
  $N_c \rightarrow \infty$ limit. \\
   
 We now wish to assign an entropy to a baryon 
 and investigate its properties. 
 We shall do this in the following way. 
 A baryon of a given spin $s$  forms an irreducible representation 
 of the flavor $SU(N)_F$-group of certain dimensionality 
 $D_F$.   
    The total number of degenerate baryonic 
 states is thus $n_{st} =(2s+1)D_F$. 
 The log of this degeneracy defines the micro-state entropy,  
 \begin{equation} \label{baryonS}
  S_B =  \ln ((2s+1)D_F) \,, 
 \end{equation} 
 which we are going to assign to the baryon. 
   As in the monopole example, we can super-softly break the flavor symmetry  
group, say, by adding the tiny quark masses.  Such a  super-soft breaking lifts the degeneracy of the baryon spectrum in a controllable way
without substantially modifying the number of relevant states. 
We can count their number and then take back the limit of the exact flavor symmetry.  \\
 
 From our previous experience, we expect that the theory should give a clear signal 
 in terms of some coupling going strong when the above  
 entropy violates the Bekenstein bound (\ref{Bek1}).  
  In order to saturate the bound, we first take the scaling $N \sim N_c$ \footnote{This regime was referred to as topological expansion of QCD in
 \cite{Veneziano}.}.  In this limit the baryon entropy scales 
 as $N$. For example, the baryon of a spin $s= {N_c \over 2}$ has 
 degeneracy
  \begin{equation}\label{NstatesB} 
 n_{st} = (N_c + 1) \begin{pmatrix}
    N_c + N -1   \\
     N_c   
\end{pmatrix} \,. 
  \end{equation}
Notice a clear similarity with the micro-state degeneracy of the monopole  (\ref{Nstates}) under the identification $N_{\sigma} \rightarrow  N_c$.    
For $N_c \sim N$ the above expression scales exponentially 
with $N$. In particular, for $N_c = N $ it is of order  
 $n_{st} \sim \sqrt{N} 2^{2N}$. 
 Thus, the baryon entropy at the boundary of the strong 't Hooft coupling, 
 \begin{equation} \label{tlimit}
   g^2N_c \sim g^2N \sim 1\,,
 \end{equation}
 is $S_B \sim N$.  Now, using (\ref{Bmass}), we obtain   
  \begin{equation} \label{AreaB} 
  S_B = N = M_BR_B  = (R_B f_{\pi})^2 \, .
 \end{equation} 
The last expression represents the same black-hole-type area law as we encountered in the theory 
with 't Hooft-Polyakov monopole in (\ref{Nsaturation}). 

\section{Confinement from Entropy Bound?} 

In the light of the above observation it is tempting to ask whether there 
 is any relation between the Bekenstein entropy bound and confinement. 
  More precisely: \\
  
  {\it  Can confinement be understood as a prevention mechanism against violation of the 
 Bekenstein entropy bound? } \\
 
  In order to make the logic of this question clearer,  
 consider the following example. 
Take a minimal case of QCD with a single quark flavor. 
Due to confinement, baryons are color-singlets.  So the degeneracy is coming  
from the spin structure. For example, a baryon of spin $s = {N_c \over 2}$ 
has entropy  $S_B = \ln(N_c +1)$. At the same time its mass should 
scale as $N_c$ in units of its inverse size. So, the maximal entropy 
allowed by the Bekenstein bound (\ref{Bek1}) is of order $S_{max} \sim N_c$. 
This is much higher than the actual entropy of the baryon
$S_B$ coming from the spin degeneracy. 
Thus, in the confining version of the theory the
baryons do not violate the Bekenstein bound,  unless, as previously discussed, we increase the number of flavors beyond the limit of  perturbative unitarity.   \\

Now, imagine that the same theory is non-confining.   
Could in this case the entropy bound be violated? Obviously, we are imagining a regime that most likely makes no physical sense.  However, we have no choice;  we wish to understand whether one of the reasons for inconsistency of such a regime is the violation of the entropy bound. \\
 
 Therefore, imagine a situation in which the colored asymptotic states are permitted. For example, consider a free quark
 of mass $M_q$, localized within the radius $R_q$ equal to its Compton wavelength  $R_q = 1/M_q$.  Its maximal entropy permitted by the Bekenstein bound (\ref{Bek1}) is $S_{max} \sim M_qR_q \sim 1$.  
On the other hand, the quark could exist in $n_{st} = 2N_c$ 
distinct degenerate spin and color states.  Therefore, we must assign to it an informational entropy
 $S_q = \ln(2N_c)$. Hence, at first glance an unconfined quark would violate Bekenstein entropy bound for sufficiently large $N_c$. \\
  
  Of course,  one has to be more careful since the wave-packets
 spread rapidly. Nevertheless, quarks could form the baryon-like multi-quark bound-states that spread much slower. However, in contrast 
 with the confining theory, in the non-confining version the bound-states could carry color and therefore be highly degenerate. 
Hence, the resulting micro-state entropy could violate the  Bekenstein bound. \\

  For example, imagine a bound-state of size $R_B$ composed out of 
  $N_q$ quarks that transforms as rank-$N_q$ antisymmetric tensor under  
  $SU(N_c)$.   For $N_c \gg N_q \gg 1$ its entropy scales as 
  $S_B \sim N_q  \ln (N_c/N_q)$.  The mass of the bound-state is expected to scale as $M_B \sim N_q/R_B$. If so,  the Bekenstein bound on its entropy 
  is expected to be $S_{max} \sim N_q$. This is smaller than the actual entropy $S_B$ due to color degeneracy.  Hence, we would have,
 \begin{equation}\label{ratio}
   {S_B \over S_{max}} \sim   \ln (N_c/N_q) \gg 1 \, .
 \end{equation} 
 Thus, the entropy bound could be violated in a non-confining theory. 
 This is something that does not happen in the confining version of the theory.  
 So, can the confinement be viewed as a preventive measure against such a violation? 
 
 Of course, the above arguments indicating that unconfined theory would violate Bekenstein bound are qualitative and hand-waving. 
 However, their sole purpose is to open up a new possible view on the
 question of confinement.

 \section{Discussions} 
 
  We have shown that in non-gravitational renormalizable quantum field theories non-perturbative states such as solitons and baryons exhibit the 
following curious behaviour: 
  The Bekenstein bound on information storage (\ref{Bek1}) is saturated simultaneously   
 with the perturbative unitarity bound; at the same time the soliton (baryon)  entropy becomes equal to its area. 

  It is remarkable that at the saturation point  the relations
 among the soliton and/or baryon characteristics become identical 
  to those of a black hole. For example, the relation between 
  the soliton size 
  and  the fundamental scale of symmetry breaking $v$ is identical to the one between the
  radius of a  black hole and the Planck mass. 
 The same relation is observed between the size of the baryon  
and the pion decay constant  $f_{\pi}$ in large-$N_c$ QCD.

   In addition, in all three cases the maximal entropies
 compatible with unitarity are expressed as the inverse dimensionless quantum coupling $g^2$ of the respective theories
evaluated at the scale $1/R$.  
  That is, we have:  
  
  \begin{eqnarray} \label{3S} 
    S_{\rm monopole} & = & {1 \over g^2_{\rm gauge}}\,,  \\ \nonumber  
    S_{\rm baryon} & = & {1 \over g_{\rm qcd}^2} \,, \\ \nonumber  
    S_{\rm BH} & = & {1 \over g^2_{\rm gr}} \,. 
 \end{eqnarray}
   
   We remind the reader that the gravitational analog of 
  a dimensionless quantum coupling 
  is $g_{gr}^2 = 1/(RM_P)^2$.    Due to this, 
 it was already noticed in \cite{NP}  that the black hole entropy can be written  as the last expression in (\ref{3S}). 
  What is remarkable is that all three seemingly unrelated objects at the 
  saturation point exhibit an universal holographic behaviour \cite{Hol1}. 
 
  Another similarity is that the soliton (baryon) size at the saturation point plays the role 
  analogous to the species scale in gravity \cite{species}. \\
  
  It is likely that the observed universal behaviour is rooted in the following picture \cite{NP}:  There exists a well defined sense in which both black holes and solitons just like baryons can be viewed as composite objects, with their energies 
 predominantly coming from the constituents of the wavelength $\sim R$ and the occupation number 
 \begin{equation} \label{key}
     N = {1 \over g^2}  \,, 
\end{equation}  
where $g^2$ must be understood as the ``running" (i.e., energy scale dependent) coupling evaluated at the scale $1/R$. 
In such a composite picture, the relation (\ref{key})  represents a maximal occupation number compatible with 
perturbativity of the collective coupling $Ng^2$. 
This was referred to as {\it maximal packing}  in \cite{NP}.   \\

 A general lesson that we draw from the presented analysis is that there is a deep underlying connection 
 between the notions of a weak-coupling and perturbative unitarity 
 on the one hand and the entropy bound and the holographic form of its saturation on the other.  This connection extends the 
concept of holography far beyond gravity.
  On the practical side, applying the bounds on information to the renormalizable quantum field theories may shed an useful light on various 
 non-perturbative regimes. One idea along these lines that we have outlined 
 is to understand confinement in QCD as a preventive mechanism agains a potential violation of the Bekenstein bound.  \\

 In summary,  we have observed  a close connection between the saturations of perturbative unitarity and Bekenstein entropy bounds by the area-law form 
 in theories with magnetic monopoles and in large-$N_c$ QCD with baryons.    
  For 't Hooft-Polyakov monopole 
  we have studied the phenomenon for two alternative mechanisms 
 of entropy enhancement in which the gapless oscillators that encode 
 quantum information come from Goldstone bosons and fermion   
zero  modes respectively. The phenomenon however appears universal. 
We have obtained similar behaviour for solitons in various dimensions, 
e.g., for Nielsen-Olesen vortexes \cite{vortex}  in $2+1$.  It also appears that renormalizability of the theory is not essential since we have observed the same behaviour 
in non-renormalizable examples, such as solitons in theories  
with $d > 4$ space-time dimensions.  The reason is that even for  non-renormalizable theories 
a dimensionless effective coupling $g^2$ and the corresponding perturbative domain are well-defined.  It is inverse of this coupling that controls 
entropy at the saturation point (\ref{3S}).  

 However, in order to keep the point sharp
and  to maximally distant ourselves from gravity, in the present 
paper we have limited ourselves by the renormalizable examples 
of theories with 't Hooft-Polyakov monopoles and QCD baryons. Other
examples, including the case of non-topological solitons \cite{giaandrei}
will be discussed elsewhere.  \\

\section*{Acknowledgements}
We thank Cesar Gomez and Goran Senjanovic for discussions. 
This work was supported in part by the Humboldt Foundation under Humboldt Professorship Award.

\end{document}